\documentclass[letter]{aa}  

\usepackage{graphicx}
\usepackage{threeparttablex}
\usepackage{txfonts}
\usepackage{supertabular,booktabs}
\usepackage[version=4]{mhchem}
\usepackage{adjustbox}
\begin{document} 
\titlerunning{First detection of HNCOH$^+$ in the interstellar medium} 
\authorrunning{Rey-Montejo et al.} 

   \title{Detection of HNCOH$^+$ in the interstellar medium}

   \author{Marta Rey-Montejo\inst{1,2}, 
           Izaskun Jim\'enez-Serra\inst{1},
           Miguel Sanz-Novo\inst{3},
           Valerio Lattanzi\inst{3},
           Sven Thorwirth\inst{4},
           V{\'i}ctor M. Rivilla\inst{1},
           Shaoshan Zeng\inst{5},
           Laura Colzi\inst{1}, 
           D. San Andr\'es\inst{1,2},
           Andr\'es Meg\'ias\inst{1},
           Antonio Mart\'inez-Henares\inst{5},
           Miguel \'Angel Requena-Torres\inst{6},
           Sergio Mart{\'i}n\inst{7,8},
           Belén Tercero\inst{9,10},
           Pablo de Vicente\inst{10} 
           }

   \institute{
    Centro de Astrobiolog{\'i}a (CAB), CSIC-INTA, Ctra. de Ajalvir km 4, 28850, Torrej\'on de Ardoz, Spain.\\
   \email{mrey@cab.inta-csic.es}
   \and  
    Departamento de F{\'i}sica de la Tierra y Astrof{\'i}sica, Facultad de Ciencias F{\'i}sicas, Universidad Complutense de Madrid, 28040 Madrid, Spain.
    \and
    Center for Astrochemical Studies, Max-Planck-Institut f\"ur extraterrestrische Physik, Giessenbachstrasse 1, Garching bei Munchen, 85748, Germany.
    \and
    I. Physikalisches Institut, Universität zu Köln, Zülpicher Str. 77, 50937 Köln, Germany
    \and
    Star and Planet Formation Laboratory, RIKEN Pioneering Research Institute, RIKEN, 2–1 Hirosawa, Wako, Saitama, 351–0198, Japan.
    \and
    Department of Physics, Astronomy, and Geosciences, Towson University, Towson, MD 21252, USA.
    \and
    European Southern Observatory, Alonso de C\'ordova, 3107, Vitacura, Santiago 763-0355, Chile.
    \and
    Joint ALMA Observatory, Alonso de C\'ordova, 3107, Vitacura, Santiago 763-0355, Chile.         
    \and
    Observatorio Astronómico Nacional (OAN, IGN), C/ Alfonso XII, 3, 28014, Madrid, Spain.
    \and
    Centro de Desarrollos Tecnológicos, Observatorio de Yebes (IGN), 19141 Yebes, Guadalajara, Spain.
    }

   \date{Received September 25, 2026; accepted October 26, 2026}

 
  \abstract
   {Only five out of the $\sim$350 molecules detected in the interstellar medium (ISM) carry the isocyanate functional group (NCO), the essential substructure of peptide bonds in proteins. Among them are isocyanic acid (HNCO) and its most stable N-protonated isomer H$_2$NCO$^+$. Previous studies have predicted that the O-protonated isomer HNCOH$^+$ should be present in the ISM; yet this species has remained elusive.}
   {We search for the metastable HNCOH$^+$ towards the  Galactic Center molecular cloud G+0.693-0.027, where H$_2$NCO$^+$ has already been detected. 
   }
   {We use an ultra-deep broadband spectral survey carried out towards G+0.693-0.027 with the Yebes 40-m, IRAM 30-m and APEX telescopes covering nearly 150 GHz of  bandwidth between 31 and 275 GHz.} 
   {We report the first detection of HNCOH$^+$ in the ISM with a column density of $N$ = (9.9 $\pm$ 0.5)$\times$10$^{12}$ cm$^{-2}$, which corresponds to a molecular abundance relative to H$_2$ of (7.3$^{+1.3} _{-1.0}$)$\times$10$^{-11}$. With the measured H$_2$NCO$^+$ column density of $N$ = (2.6 $\pm$ 0.3) $\times 10^{12}$ cm$^{-2}$, this yields a HNCOH$^+$/H$_2$NCO$^+$ ratio of 3.8$^{+0.5} _{-0.4}$. This result is consistent with recent kinetic calculations that favour the formation of HNCOH$^+$ over H$_2$NCO$^+$.}
  {The detection of HNCOH$^+$ in the ISM broadens the inventory of molecules known to contain the isocyanate functional group and reinforces recent works showing that isomer abundances in the ISM do not necessarily correlate with thermodynamic stability.}


   \maketitle
%

\section{Introduction}
A substantial number of molecules detected in space can be assembled from only the four elements H, C, N, and O. These elements constitute the fundamental building blocks of organic and prebiotic chemistry and, when arranged appropriately, give rise to key functional groups such as carboxyl (-COOH) and amino (-NH$_2$), present in biologically relevant molecules such as amino acids. Alternatively, they can link together forming a peptide bond –C(=O)–N(–H)–, enabling the formation of large proteins \citep{Marcelino2018}. The detection of simple molecules containing the isocyanate group (-N=C=O) in the interstellar medium (ISM) is particularly significant, as they may offer insight into the earliest chemical steps leading to peptide synthesis.

Within the current inventory of $>$350 molecules identified in the ISM or in circumstellar envelopes\footnote{\url{https://cdms.astro.uni-koeln.de/classic/molecules}}, only a very small subset carries the isocyanate functional group. To date, this subset includes: isocyanic radical \citep[NCO;][]{Marcelino2018}; isocyanic acid \citep[HNCO;][]{Snyder1972}; N‑protonated isocyanic acid  \citep[H$_2$NCO$^+$;][]{Marcelino2018}; methyl isocyanate \citep[CH$_3$NCO;][]{Halfen2015}; and ethyl isocyanate \citep[C$_2$H$_5$NCO;][]{RodriguezAlmeida2021}.

HNCO exhibits several metastable isomers in the singlet ground electronic state: HOCN, HCNO, and HONC, which lie at 24.7, 70.7, and 84.1 kcal mol$^{-1}$ (12430, 35578, and 42321 K) above HNCO, respectively \citep{Schuurman2004}. These species can be formed through a variety of chemical pathways. In the gas phase, the protonation and subsequent hydrogenation of the radicals NCO and CNO yield the protonated [C,H,N,O] ions, including H$_2$NCO$^+$, HNCOH$^+$, HCNOH$^+$, H$_2$OCN$^+$, and H$_2$CNO$^+$ \citep{Marcelino2010, Quan2010, Quenard2018, Hochlaf2026}. By dissociative recombination (DR) reactions, these ions  produce all the [C,H,N,O] isomers, among them HONC, the highest-energy isomer, which has not yet been detected in space \citep{Marcelino2018}. 

Quantum-chemical calculations reveal that the N-protonated variant of isocyanic acid, H$_2$NCO$^+$, is the global minimum, followed by the O-protonated variant HNCOH$^+$ \citep[18 kcal mol$^{-1}$, $\sim$9060 K;][]{Lattanzi2012,Hochlaf2026}. 
Furthermore, \cite{Muramaki2024} investigated the thermal rate coefficients and branching ratios for the gas-phase reactions
H$_3 ^+$ + HNCO $\rightarrow$ H$_2$ + HNCOH$^+$/H$_2$NCO$^+$ which is expected to be one of the main pathways for forming these species, since protonation by H$_3 ^+$ in the ISM has been shown to be one of the most efficient mechanisms \citep{Oka2006}. Their calculations showed that the channel leading to HNCOH$^+$ is favoured over the formation of H$_2$NCO$^+$, with branching ratios of $\sim$80\% and $\sim$20\%, respectively. These results are consistent with laboratory rotational spectroscopy measurements, which reveal that the abundance of HNCOH$^+$ in the original Fourier-transform microwave spectroscopy (FTMW) study is twice that of H$_2$NCO$^+$ \citep{Lattanzi2012}.
Moreover, the proton-affinity-induced isomerisation from the meta-stable HNCOH$^+$ to the stable H$_2$NCO$^+$ is unlikely to occur due to the substantial high barrier \citep[100 kcal mol$^{-1}$, $\sim$50320 K;][]{Ijjaali2001} between them.
Despite this, only H$_2$NCO$^+$ has been detected in the ISM \citep{Marcelino2018,RodriguezAlmeida2021}. Therefore, the detection of  HNCOH$^+$ yields a key step that had been missing in our understanding of the interstellar chemistry of HNCO and its protonated isomers.

In this Letter, we report the first detection of HNCOH$^+$ in the ISM towards the Galactic Center molecular cloud G+0.693-0.027 (hereafter, G+0.693), a source known for its exceptionally rich chemistry driven by large‑scale shocks, likely induced by a cloud–cloud collision \citep{Zeng2020}. These shocks efficiently sputter material from both the icy mantles and refractory cores of dust grains, injecting large amounts of molecular material into the gas phase. As a result, G+0.693 stands out as one of the most prolific astronomical environments for identifying new interstellar molecules.

\section{Observations}\label{sec:obs}
We have used an ultra-deep broadband spectral survey obtained towards the molecular cloud G+0.693 at 7 mm, 3 mm, 2 mm, and 1 mm with the Yebes 40-m, IRAM 30-m, and APEX radio telescopes.
The Yebes 40-m observations were conducted using the  ultra-broadband Nanocosmos Q-band (7 mm) HEMT receiver to cover the entire Q-band frequency range $\sim$31.1-50.4 GHz (projects 20A008 and 21A014). The IRAM 30-m observations were carried out using the Eight MIxer Receiver (EMIR) and the Fast Fourier Transform Spectrometer (FFTS) with 200 kHz of resolution across the spectral ranges 71.8–116.7 GHz, 124.77–175.5 GHz, 199.8–238.29 GHz, 252.52–260.30 GHz, and 268.2–275.98 GHz (projects 172-18, 018-19, 133-19 and 123-22), providing complementary coverage. For the APEX observations, the NFLASH230 receiver was connected to two FFTS backends, which provided a simultaneous coverage of two sidebands of 7.9 GHz each separated by 8 GHz. The covered spectral ranges were 217.9-225.9, 234.18-242.18, 243.94-252.12, and 260.18-268.37 GHz \citep[see][project E-0108.C-0306A-2021]{Rivilla2022}. 
Data were collected in position-switching mode, centered at $\alpha = 17^{\rm h}47^{\rm m}22^{\rm s}$, $\delta=-28^{\circ} 21\arcmin 27\arcsec$, with the off position shifted by $\Delta\alpha=-885\arcsec$ and $\Delta\delta=290\arcsec$.

The spectra of Yebes, IRAM and APEX observations were finally smoothed to a frequency resolution of 256 kHz (1.5-2.5 km s$^{-1}$), 615 kHz (1.0-2.2 km s$^{-1}$) and 1 MHz (1.1-1.4 km s$^{-1}$), respectively. 
The half power beam width (HPBW) was $\sim$35$\arcsec$-55$\arcsec$ for Yebes 40-m, $\sim$14$\arcsec$-29$\arcsec$ for IRAM 30-m, and $\sim$23$\arcsec$-28$\arcsec$ for APEX in the observed frequency ranges. Note, however, that the molecular emission towards G+0.693 is extended over the beam of the three telescopes \citep{Li2020, Zheng2024} and hence, the measured spectra are given in units of antenna temperature ($T_{\rm A}^*$). We refer to \citet{Rivilla2022,Rivilla2023} and \citet{SanzNovo2023} for further information on all observations and data reduction. 

\section{Detection of HNCOH$^+$}
The astronomical line identification of HNCOH$^+$ has been accomplished using frequency predictions based on the reported laboratory microwave spectrum observed by \cite{Lattanzi2012}. As only the two lowest a-type rotational transitions 1$_{0,1}$-0$_{0,0}$ and 2$_{0,2}$-1$_{0,1}$ at about 20 and 40 GHz have been studied in the laboratory so far \citep{Lattanzi2012}, the spectroscopic entry was extrapolated to 300 GHz using SPCAT, and then incorporated into \textsc{Madcuba} package (version 12.1.1)\footnote{\url{https://cab.inta-csic.es/madcuba/}}. 
The identification of the molecular lines was carried out using the Spectral Line Identification and Modelling (\textsc{Slim}) tool from \textsc{Madcuba} \citep{Martin2019}, which assumes local thermodynamic equilibrium (LTE) to compute the synthetic spectra. To derive the physical parameters of the emission, we used the \textsc{Autofit} tool of \textsc{Slim} that provides a nonlinear least-squares LTE fit to the data using the Levenberg-Marquardt algorithm. The fit provides information about the column density, $N$, excitation temperature, $T_{\rm ex}$, radial velocity, $v_{\rm LSR}$, and line width, FWHM, of the molecular emission. We note that owing to the typically broad FWHM of the molecular emission observed in G+0.693 \citep[line width of $\sim$15-22 km s$^{-1}$; e.g.][]{Izaskun2020, Rivilla2022b}, the quadrupole hyperfine structure from the presence of the $^{14}$N (I=1) nucleus is not resolved but collapsed to just one line.

Eight a-type rotational transitions (i.e., with $\Delta K_a$ = 0 and $\Delta K_c$ = $\pm$ 1; $\mu _a$ = 1.3 D) of HNCOH$^+$ are covered by the observational survey, spanning much of the $J' \rightarrow J''$ progression with $J'$ = 2 to 13. However, note that the predicted frequencies for the transitions with $J'$ = 3, 6, 9 and 10 fall within the gaps of the observed frequency coverage of our survey (see Appendix~\ref{sec:spectral_cov}). Moreover, the 13$_{0,13}$-12$_{0,12}$ transition at 258.9 GHz was excluded from the analysis due to blending with an atmospheric line.

\begin{figure*}[h]
    \centering
    \includegraphics[width=\linewidth]{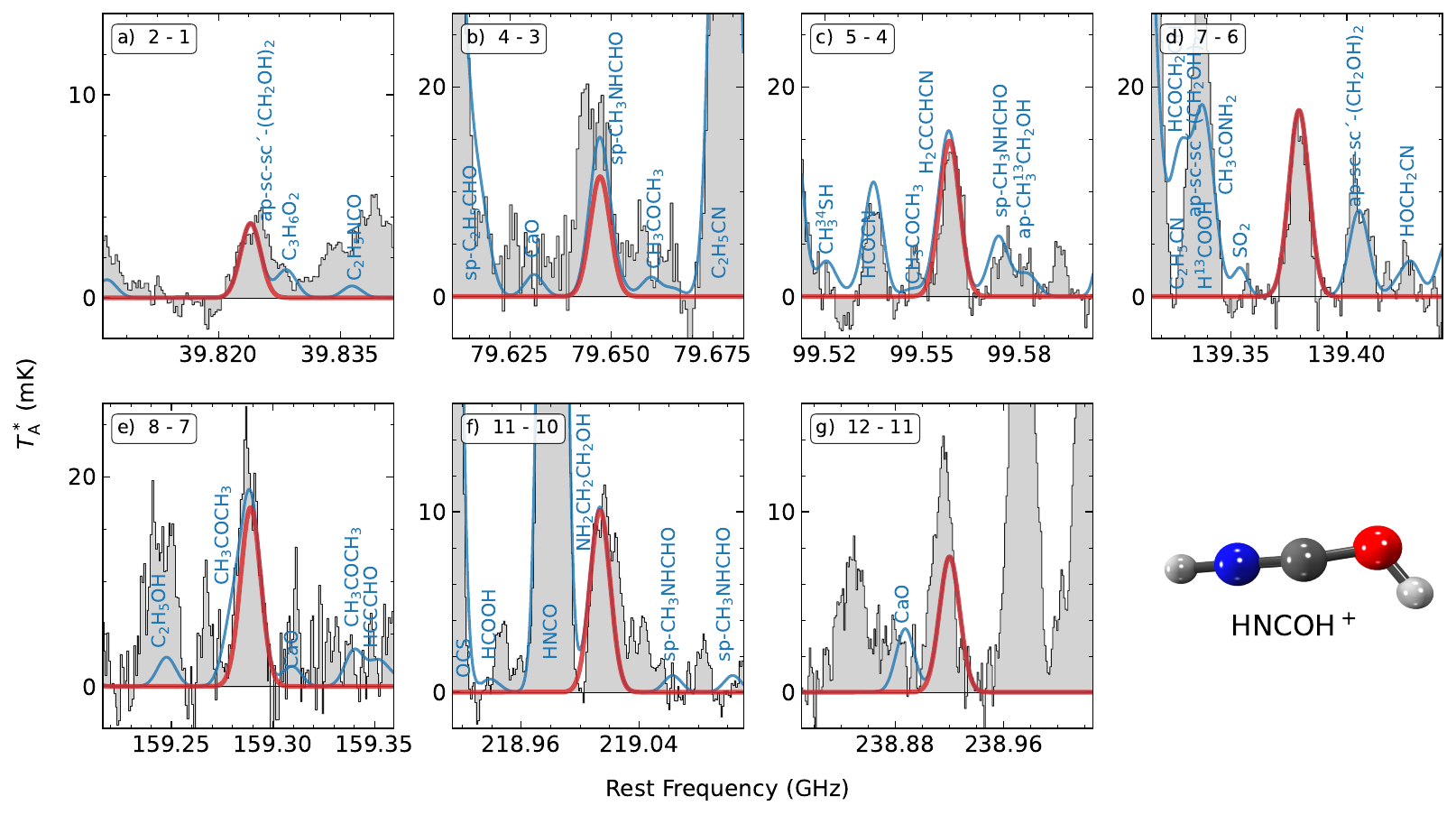}
    \caption{$K_a$ = 0 rotational transitions of HNCOH$^+$ identified towards G+0.693 and sorted by increasing frequency (see also Table~\ref{tab:spectroscopic_HNCOH+}). The LTE fit of HNCOH$^+$ is plotted with a red line, while the blue line depicts the emission from all the molecules identified to date in our survey, including HNCOH$^+$ (observed spectra shown as gray histograms). The molecular structure of HNCOH$^+$ is shown in the bottom right. The quantum numbers of each rotational transition ($J$ + 1 $\rightarrow$ $J$) are indicated in the upper left of each panel. Blue labels show the rest of identified species contributing to the observed spectra.
    }
    \label{fig:HNCOHp}
\end{figure*}

The detected transitions are shown in Figure~\ref{fig:HNCOHp} and their spectroscopic information is listed in Table~\ref{tab:spectroscopic_HNCOH+}.  
The best LTE fit provided a value of FWHM = 21.9 $\pm$ 1.4 km s$^{-1}$, 
and a velocity of $v_{\rm LSR}$ = 66.9 $\pm$ 0.6 km s$^{-1}$, consistent with the values found in G+0.693 for other molecular species, specially molecular ions \citep[see e.g.][]{SanzNovo2024}. The excitation temperature derived is $T_{\rm ex}$ = 24.4 $\pm$ 1.3 K, while the derived column density is $N$ = (9.9 $\pm$ 0.5)$\times$10$^{12}$ cm$^{-2}$. Assuming an H$_2$ column density of $N_{\rm H_2}$ = (1.35 $\pm$ 0.20)$\times$10$^{23}$ cm$^{-2}$ inferred from C$^{18}$O emission at the velocity component of G+0.693 \citep{Sergio2008}, the HNCOH$^+$ column density translates into a molecular abundance with respect to H$_2$ of (7.3$^{+1.3} _{-1.0}$)$\times$10$^{-11}$ towards G+0.693. As shown in Table~\ref{tab:spectroscopic_HNCOH+}, the detected transitions cover a wide range of upper-level energies, $E_{\rm up}$, which allowed us to derive an accurate estimate on $T_{\rm ex}$. The consistency across these features, together with the detection of a nearly complete $K_a$ = 0 rotational progression within the spectral coverage of the G+0.693 survey, yields a robust identification of HNCOH$^+$.

The astronomical lines observed here, except transition 2$_{0,2}$-1$_{0,1}$, represent a first record of the millimetre-wave spectrum, identified in space towards G+0.693 and not measured before in the laboratory. This implies that, although the uncertainties can be as high as 1.2 MHz at 238 GHz, which translates into velocity uncertainties between 1.0-1.5 km s$^{-1}$, these would not have an impact on the analysis as they are small compared to the typical line widths of the molecular line emission measured towards G+0.693 (between 15-22 km s$^{-1}$). An analogous extrapolation approach was used in the interstellar discovery of HOCS$^+$ \citep{Miguel2024a}, and then corroborated in the laboratory by \cite{Lattanzi2024}.
Furthermore, the inclusion of the isolated 5$_{0,5}$–4$_{0,4}$, 7$_{0,7}$–6$_{0,6}$, and 8$_{0,8}$–7$_{0,7}$ transitions together with the microwave data set by \cite{Lattanzi2012} provides a first estimate of the centrifugal distortion constant $D_J$ of (2.61 $\pm$ 0.13)$\times10^{-3}$ MHz. This value is in very good agreement with the value of 2.83$\times10^{-3}$ MHz obtained from the harmonic force field calculated at the CCSD(T)/cc-pVQZ level of theory \citep{Lattanzi2012}. Nevertheless, a comprehensive description of the pure rotational spectrum still requires a thorough targeted laboratory study.

We have also performed a complementary rotational diagram analysis (Goldsmith \& Langer 1999) with \textsc{Madcuba}, using the transitions shown in Figure~\ref{fig:HNCOHp}.  As expected, given the optically thin emission of the observed transitions, the derived physical parameters of HNCOH$^+$ are consistent with the ones obtained from the \textsc{Autofit} analysis: $N$ = (9.2 $\pm$ 1.4)$\times$10$^{12}$ cm$^{-2}$ and $T_{\rm ex}$ = 23.3 $\pm$ 1.6 K (Figure~\ref{fig:rot_HNCOHp}).

\section{Discussion}
To understand the implications of finding interstellar HNCOH$^+$ in the chemistry of HNCO, we compare the abundance of HNCOH$^+$ with H$_2$NCO$^+$. Due to the availability of new, higher-sensitivity data for the studied source as compared to those of \citet{RodriguezAlmeida2021}, we have recalculated the LTE fit of H$_2$NCO$^+$ for G+0.693 using the same transitions reported by \cite{RodriguezAlmeida2021}, which yield a column density of $N$ = (2.6 $\pm$ 0.3) $\times 10^{12}$ cm$^{-2}$, consistent with previous results within a factor of 1.5 (see Appendix~\ref{sec:H2NCOp}).
Thus, we find a HNCOH$^+$/H$_2$NCO$^+$ ratio of 3.8$^{+0.5} _{-0.4}$ towards G+0.693. 
The quantum-chemical calculations employed by \cite{Lattanzi2012} show that the H$_2$NCO$^+$ isomer is the most favourable product from a thermodynamic point of view, with a difference in energy of 18 kcal mol$^{-1}$ ($\sim$9060 K). Following the minimum energy principle (MEP) formulation \citep{Lattelais2009}, the abundance ratios between structural isomers must be governed by thermodynamic equilibrium. Therefore, the abundance of HNCOH$^+$ based on thermodynamical considerations is predicted to be more than 20 orders of magnitude lower in comparison to H$_2$NCO$^+$ at the $T_{\rm kin}$ of G+0.693 \citep[$\sim$70–140 K;][]{Zeng2018}. Thus, the derived abundance of HNCOH$^+$ must be explained in terms of kinetics instead.
The theoretical kinetic results derived by \cite{Muramaki2024} indicate that the gas-phase reactions H$_3$$^+$ + HNCO $\rightarrow$ H$_2$ + HNCOH$^+$/H$_2$NCO$^+$ preferentially proceed via the mechanism leading to the less stable isomer, HNCOH$^+$, rather than to the more thermodynamically favourable product H$_2$NCO$^+$.
This result does not emerge as the unique exception to the MEP, since recent detections in the ISM of many other isomeric families have demonstrated that high-energy structural isomers can share similar abundances to their more stable counterparts \citep[e.g.][]{Shingledecker2019,Juan2021,SanAndres2024, Zeng2025, Rivilla2026}. 

Furthermore, we compared the abundance ratios of HNCOH$^+$ and H$_2$NCO$^+$ over HNCO \citep{Zeng2018}. The HNCOH$^+$/HNCO ratio is (2.95 $\pm$ 0.17)$\times 10^{-3}$, whereas the H$_2$NCO$^+$/HNCO is (7.7 $\pm$ 0.9)$\times 10^{-4}$. Taking into account that the proton affinity of HNCO is 759 KJ mol$^{-1}$ \citep{Wight1980}, this result is consistent with the abundance ratios found in cold dense clouds for neutral molecules with proton aﬃnities above 700 kJ mol$^{-1}$ \citep{Agundez2022}.

Moreover, it is worth noting that the $T_{\rm ex}$ of HNCOH$^+$ is higher than the one found for H$_2$NCO$^+$ ($T_{\rm ex}$ of $\sim$24 K vs. $\sim$9 K, respectively), which is consistent with a similar trend observed in $T_{\rm ex}$ for other O-protonated cations in G+0.693 \citep{SanzNovo2024}. A possible explanation for this lies in the different dipole moments. Molecules with larger dipole moments have correspondingly larger Einstein coefficients ($A_{ij}$), which imply higher critical densities. This is expected to yield lower $T_{\rm ex}$ because the emission is likely sub-thermally excited given that the H$_2$ gas densities in G+0.693 are of a few 10$^{4}$ cm$^{-3}$ \citep{colzi24}. The dipole moment of H$_2$NCO$^+$ is $\mu _a$ = 4.1 D, whereas the dipole moment components of HNCOH$^+$ are $\mu _a$ = 1.3 D and $\mu _b$ = 1.7 D \citep{Lattanzi2012}. Therefore, the higher $T_{\rm ex}$ value found for HNCOH$^+$ is consistent with its lower dipole moment compared to H$_2$NCO$^+$.

\section{Conclusion}
We report the first detection of HNCOH$^+$ in the ISM towards the Galactic Center molecular cloud G+0.693. The derived molecular abundance is (7.3$^{+1.3}_{-1.0}$)$\times$10$^{-11}$ with respect to H$_2$. This species represents the sixth identified molecule containing the isocyanate functional group, the backbone structure of peptides. Moreover, the detection reveals that the fractional abundance of HNCOH$^+$ is approximately four times larger than that of H$_2$NCO$^+$, the most stable isomer of isocyanic acid. This finding demonstrates that, within the ISM, the thermodynamically most stable isomers are not necessarily the most abundant ones.

As the pioneering microwave laboratory study of \cite{Lattanzi2012} covered only the two lowest rotational transitions of HNCOH$^+$, the present study provides first insight into the millimetre-wave spectrum. The $D_J$ centrifugal distortion parameter, derived by considering three virtually unblended transitions (Table~\ref{tab:spectroscopic_HNCOH+}), is in near-perfect agreement with the value derived from high-level force-field calculations. A more detailed characterization of the rotational spectrum, including transitions with $K_a$ $>$ 0, will require dedicated laboratory studies in the future.

\begin{acknowledgements}
    This work is supported by ERC grant OPENS, GA No. 101125858, funded by the European Union. Views and opinions expressed are however those of the author(s) only and do not necessarily reflect those of the European Union or the European Research Council Executive Agency. Neither the European Union nor the granting authority can be held responsible for them. M.R.-M, I.J.-S. and A.M. acknowledge funding from the Spanish National Research Council (CSIC) through the Bilateral project SOULMATE (BIJSP25017). I.J-.S, V.M.R, L.C, M.S.-N, A.M. and D.S.A acknowledge funding from grant PID2022-136814NB-I00 funded by the Spanish Ministry of Science, Innovation and Universities / State Agency of Research MICIU/AEI/10.13039/501100011033 and by “ERDF/EU”.
    M.S.-N. acknowledges funding from the Marie Sklodowska Curie fellowship SOUL (project number 101272259) funded by the European Union.
    V.M.R, M.S.-N and D.S.A also aknowledge the grant CNS2023-144464 funded by MICIU/AEI/10.13039/501100011033 and by “European Union NextGenerationEU/PRTR”.
    D.S.A. thanks the financial support provided by the Comunidad de Madrid through the Grant PIPF-2022/TEC-25475.
    D.S.A. and V.M.R. acknowledge the funds provided by the Consejo Superior de Investigaciones Cient{\'i}ficas (CSIC) and the Centro de Astrobiolog{\'i}a (CAB) through the project 20225AT015 (Proyectos intramurales especiales del CSIC). 
    S.Z. and A.M.-H. acknowledges the support by RIKEN Special Postdoctoral Researchers Program.
    B.T. thanks projects PID2022-137980NB-I00 and PID2023-147545NB-I00 funded by the Spanish Ministry of Science and Innovation/State Agency of Research MCIN/AEI/10.13039/501100011033 and by “ERDF A way of making Europe".
    At Cologne, this work has been supported by an ERC Advanced Grant (MissIons: 101020583) and by the Deutsche Forschungsgemeinschaft (DFG) via the Collaborative Research Centre 1601 (project ID: 500700252, subprojects A4 and B8).
    L.C. acknowledges support of a fellowship from the “la Caixa” Foundation (ID 100010434). The fellowship code is LCF/BQ/PR25/12110012.
\end{acknowledgements}

\bibliographystyle{aa}
\bibliography{report}

\begin{appendix} 
\label{sec:appendix}
\renewcommand{\arraystretch}{1.2} 
\section{Spectral coverage of the observations}\label{sec:spectral_cov}
In the following table we show the central frequency of the rotational ($J$) level progression in the $K_a$ = 0 ladder for HNCOH$^+$. For a-type transitions in the R-branch (i.e. $J'$ = $J''$ + 1), the line frequencies are approximately distributed according to ($B + C$)($J''$ + 1), where $B$ and $C$ are the rotational constants. Using the value ($B + C$) = 19.9182 GHz reported by \cite{Lattanzi2012}, we computed the expected central frequency for each transition in the $K_a$ = 0 ladder (Table~\ref{tab:spectral_coverage}).
It can clearly be seen that the missing $J' \rightarrow J''$ transitions in the analysis correspond to the gaps in the observational data. However, it is worth noting that 3 of the missing transitions lie in the absorption bands of the atmosphere (marked with an asterisk in Table~\ref{tab:spectral_coverage}) and therefore, cannot be observed with ground-based radio telescopes.
\begin{table}[h]
\begin{center}
    \begin{ThreePartTable}
    \caption{Spectral coverage of the $K_a$ = 0 ladder for HNCOH$^+$.}
        \begin{tabular}{c c c}
        \toprule
        \toprule
        $J'\rightarrow J''$ & Frequency (GHz) & Telescope\\
        \midrule
        2 - 1 & 39.8 & Yebes 40-m\\
        3 - 2 & 59.7 & * \\
        4 - 3 & 79.6 & IRAM 30-m\\
        5 - 4 & 99.5 & IRAM 30-m\\
        6 - 5 & 119.5 & * \\
        7 - 6 & 139.4 & IRAM 30-m\\
        8 - 7 & 159.3 & IRAM 30-m\\
        9 - 8 & 179.3 & *\\
        10 - 9 & 199.2 & -\\
        11 - 10 & 219.1 & APEX\\
        12 - 11 & 239.0 & APEX\\
        13 - 12 & 258.9 & IRAM 30-m\\
        \bottomrule
        \end{tabular}
        \label{tab:spectral_coverage} 
    \end{ThreePartTable}
    \begin{tablenotes}
        NOTE. The `-' symbol corresponds to the gaps in the observational data, whereas the `*' symbol corresponds to transitions that lay in the atmospheric absorption bands, which were not covered either in our observations.
    \end{tablenotes}
\end{center}
\end{table}

\section{Spectroscopic information of HNCOH$^+$ detection.}\label{sec:spectral_table}

In Table \ref{tab:spectroscopic_HNCOH+}, we report the rotational transitions of HNCOH$^+$ computed from the spectroscopic constants of \cite{Lattanzi2012} and considered in the present study. We note that the $^{14}$N nuclear quadrupole hyperfine structure has been omitted, as it cannot be resolved owing to the broad line widths of the molecular emission observed in G+0.693. All transition frequencies but that of the 2$_{0,2}$-1$_{0,1}$ transition are extrapolated based on the data given in \cite{Lattanzi2012}, for which the uncertainties can be as high as 1.2 MHz at 238 GHz, which translates into velocity uncertainties between 1.0-1.5 km s$^{-1}$. These uncertainties are however small compared to the line widths of 15-22 km s$^{-1}$ measured for the molecular emission toward G+0.693 \citep{Zeng2018, Izaskun2020}.

To identify the unblended transitions, we adopt the criteria used in recent molecular detections towards G+0.693 \citep{Izaskun2026}. A spectral feature is considered predominantly unblended when its positive residual contribution\footnote{The residuals are computed subtracting the area underneath the LTE fit of the HNCOH$^+$ lines to the total measured area in the observed spectra.} (reported as residual excess in Table~\ref{tab:spectroscopic_HNCOH+}) over the velocity interval $v_{\rm LSR}$ = 66.9 $\pm$ FWHM km s$^{-1}$, is $\leq$25\%\footnote{Note that a velocity interval of $\pm$FWHM encompasses 98.1\% of the total area underneath a Gaussian line profile \citep{Izaskun2026}}. 
Following this criterion, we identify 4 unblended spectral features (see Table~\ref{tab:spectroscopic_HNCOH+}). However, if we account for the emission of the species already detected towards G+0.693 (reported as "Residual All species" in Table~\ref{tab:spectroscopic_HNCOH+}), we find that 5 features blended with known molecules are well reproduced in the observed spectra, with a contribution on the residual of $\leq$25\%. Note that features blended with only unidentified species (U-lines in Table~\ref{tab:spectroscopic_HNCOH+}) remain blended under this criterion, although they are fully consistent with the observed spectra.

\begin{table*}[h]
\begin{center}
\begin{ThreePartTable}
\caption{List of observed transitions of HNCOH$^+$ in order of increasing frequency.}
\setlength{\tabcolsep}{2pt}

        \begin{tabular}{c c c c c c c c c c}
        \toprule
        \toprule
        Frequency & Transition$^a$ & log $I$ & $E_{\rm u}$ & rms & $\int T_A ^* {\rm d}v$ & S/N$^b$ & Residual & Residual & Blending\\
        (GHz) & ($J'_{K'_a,K'_c}- J''_{K''_a,K''_c}$) & ($\rm nm^{2}~MHz$) & (K) & (mK) & (mK km s$^{-1}$) & & excess & All species & \\
        \midrule
        39.8236960 & $2_{0,2}-1_{0,1}$ & -4.8525 & 2.6 & 0.37 & 86 & 16 & 26\% & 25\% & U-line + aGg'-(CH$_2$OH)$_2$\\
        \midrule
        79.6467092 & $4_{0,4}-3_{0,3}$ & -3.9578 & 7.8 & 1.9 & 268 & 14 & 54\% & 39\% & sp-CH$_3$NHCHO + U-line\\
        \midrule
        99.5577465 & $5_{0,5}-4_{0,4}$ & -3.6733 & 11.4 & 1.2 & 347 & 32 & 8\% & 7\% & Unblended 
        \\
        \midrule
        139.3784555 & $7_{0,7}-6_{0,6}$ & -3.2515 & 20.6 & 1.3 & 416 & 42 & 0.8\% & 0.8\% & Unblended\\
        \midrule
        159.2879566 & $8_{0,8}-7_{0,7}$ & -3.0879 & 26.2 & 2.1 & 399 & 27 & 25\% & 11\% & CH$_3$COCH$_3$ \\
        \midrule
        219.0120224 & $11_{0,11}-10_{0,10}$ & -2.7124 & 47.0 & 1.2 & 236 & 32 & 18\% & 17\% & U-line + NH$_2$CH$_2$CH$_2$OH\\
        \midrule
        238.9182808 & $12_{0,12}-11_{0,11}$ & -2.6150 & 55.3 & 1.3 & 176 & 23 & 47\% & 47\% & U-line\\
        \bottomrule
        \end{tabular}

        \begin{tablenotes}
            \item NOTE. We provide the transition frequencies, quantum numbers, base 10 logarithms of the integrated intensity at 300 K (log $I$), energies of the upper levels (E$_{\rm u}$), noise level (rms), integrated area ($\int T_A ^* {\rm d}v$), signal to noise ratio (S/N) calculated with the integrated intensity and residual of each transition. The last column indicates the line blending with other molecular species. 
            \footnotesize
            \item[a] {The rotational levels of HNCOH$^+$ are expressed in terms of three quantum numbers $J_{K_a,K_c}$: $J$, the rotational quantum number; $K_a$ and $K_c$, which are the projections of $J$ onto the A and C molecular axes, respectively.}
            \item [b]{The S/N is calculated from the integrated signal ($\int T_A ^* {\rm d}v$) and the uncertainty derived as $\sigma= \rm rms\times\sqrt{\delta v \times \rm 2\times FWHM}$, where $\delta v$ is the velocity resolution of the spectra, and FWHM the line width fitted from the data.}
        \end{tablenotes}

\label{tab:spectroscopic_HNCOH+}
\end{ThreePartTable}
\end{center}
\end{table*}

\section{Rotational diagram}\label{sec:rot_diag}
\begin{figure}[h]
    \centering
    \includegraphics[width=\linewidth]{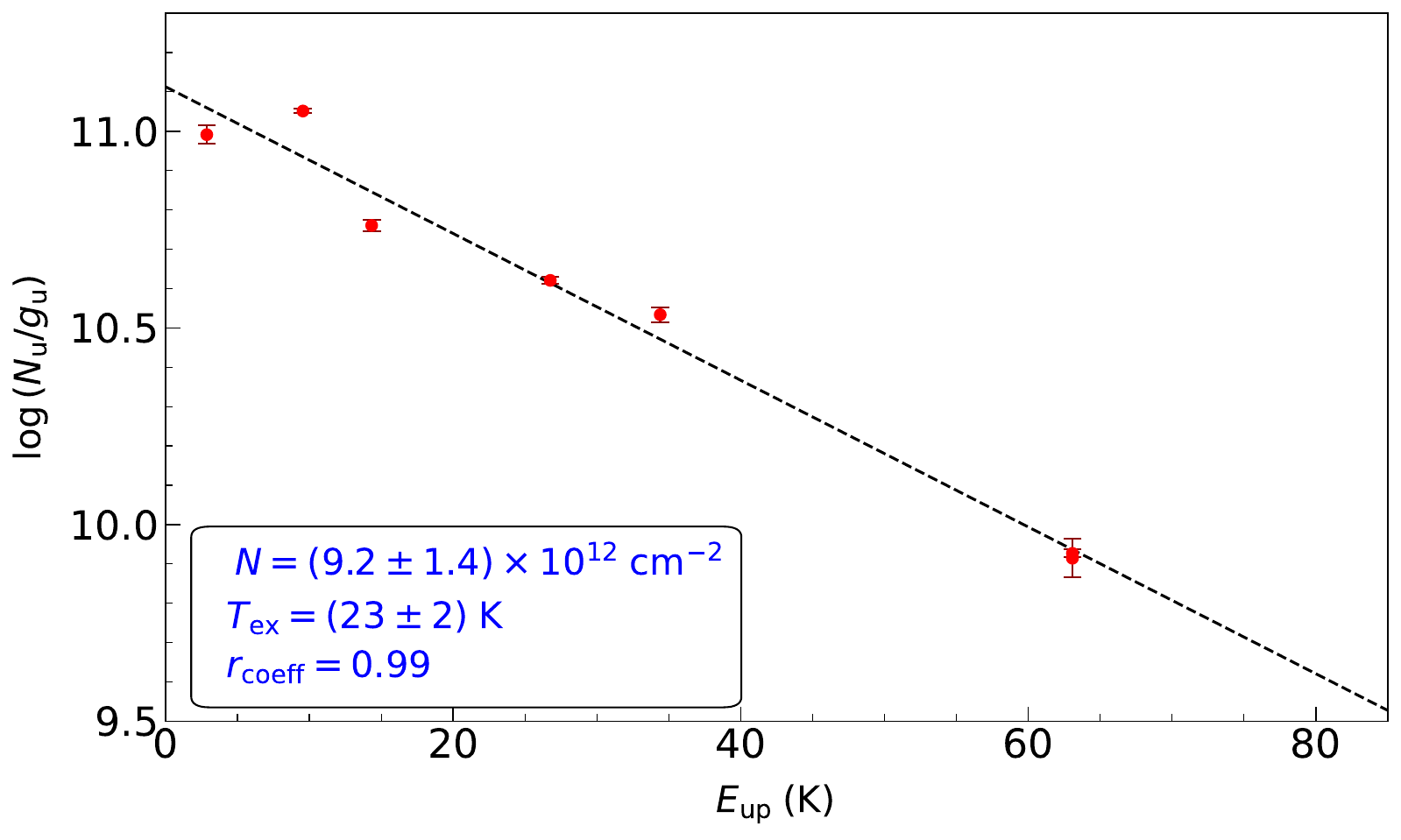}
    \caption{Rotation diagram of HNCOH$^+$ towards G+0.693 (red dots). The black dashed line corresponds to the best linear fit to the data points. The derived values for the molecular column density, $N$, the excitation temperature, $T_{\rm ex}$, and the fitting correlation coefficient, $r_{\rm coeff}$, are shown in blue.}
    \label{fig:rot_HNCOHp}
\end{figure}

We have carried out a rotational diagram analysis (Figure~\ref{fig:rot_HNCOHp}) of the emission of HNCOH$^+$ with \textsc{Madcuba} excluding transition 12$_{0,12}$-11$_{0,11}$ at 238.91 GHz due to heavy blending with unknown species. The calculation was done considering the observed velocity-integrated intensity over the line width \citep{Rivilla2021}. As described in \cite{SanAndres2024}, \textsc{Madcuba}'s rotational diagram also considers the predicted emission from all of the molecular species previously detected in the source, and accounts for the blending in each transition by subtracting the predicted line contribution of the blending species from the observed spectrum.

The derived physical parameters of HNCOH$^+$ from the rotational diagram are: $N$ = (9.2 $\pm$ 1.4)$\times$10$^{12}$ cm$^{-2}$ and $T_{\rm ex}$ = 23.3 $\pm$ 1.6 K.

\section{Re-analysis of H$_2$NCO$^+$ towards G+0.693}\label{sec:H2NCOp}
We have revised the previously detected H$_2$NCO$^+$ towards G+0.693. Following the approach carried out by \cite{RodriguezAlmeida2021} for H$_2$NCO$^+$, we carry out the analysis for the same transitions reported in that work, separating the ortho ($K_a$ = odd) and para ($K_a$ = even) states \citep[energy difference of $\sim$15 K;][]{Gupta2013}.
We used the \textsc{Madcuba} script reported in \cite{Zeng2025} to perform the $K_a$ ladder separation analysis. It first separates the $K_a$ ladder automatically, and then, for each $K_a$ ladder, the script corrects the energy of the rotational levels by subtracting the energy of the lower-state energy level. Finally, it recalculates the partition function as a direct summation of all the energy levels within each ladder. The \textsc{Autofit} tool was then applied to obtain the best-fitting LTE model for the transitions of each $K_a$ ladder, and the total column density was derived as a sum of the column density of each ladder.

For the o-H$_2$NCO$^+$ state we obtain 
$N$ = (1.7 $\pm$ 0.3) $\times 10^{12}$ cm$^{-2}$, $T_{\rm ex}$ = 8.6 $\pm$ 1.5 K, with fixed value of $v_{\rm LSR}$ = 69 km s$^{-1}$ and FWHM = 18 km s$^{-1}$ to allow the convergence. For the para-state, we assumed the same $T_{\rm ex}$ and obtained $N$ = (8.6 $\pm$ 1.4) $\times 10^{11}$ cm$^{-2}$ with fixed $v_{\rm LSR}$ = 69 km s$^{-1}$ and FWHM = 18 km s$^{-1}$.

These results yield a total column density of $N$ = (2.6 $\pm$ 0.3) $\times 10^{12}$ cm$^{-2}$ for H$_2$NCO$^+$, derived from the summed contributions of the para and ortho states, which translates into a molecular abundance with respect to H$_2$ of (1.9$^{+0.4}_{-0.3}$) $\times 10^{-11}$.  This is consistent with the column density derived by \cite[][$N$ = (1.7 $\pm$ 0.2) $\times 10^{12}$ cm$^{-2}$]{RodriguezAlmeida2021} within a factor of 1.5 . Note that the difference arises from the recalculation of the partition function taking into account the correction of the energy of the rotational levels for the separation of the $K_a$ ladder as well as the new astronomical data. This yields a slightly higher $T_{\rm ex}$ compared to that of \cite{RodriguezAlmeida2021}, but consistent within the error with the value derived in that work.

Note that the associated uncertainty for all calculations have been calculated using the Python library RichValues\footnote{\url{https://pypi.org/project/richvalues}} \citep{Andres2025}, and arise from the error propagation of the observational data.

\end{appendix}

\end{document}